\documentstyle[aps,psfig]{revtex}
\begin{document}
\draft
\renewcommand{\thefootnote}{\fnsymbol{footnote}}
\begin{title}
{\bf Numerical study of metastability due to tunneling: \\
The quantum string method}
\end{title}
\author{Tiezheng Qian}
\address{Department of Mathematics,
Hong Kong University of Science and Technology,\\
Clear Water Bay, Kowloon, Hong Kong, China}
\author{Weiqing Ren}
\address
{Department of Mathematics, Princeton University,\\
Princeton, New Jersey 08544, USA}
\author{Jing Shi}
\address
{Program in Applied and Computational Mathematics, Princeton University,\\
Princeton, New Jersey 08544, USA}
\author{Weinan E}
\address{Department of Mathematics and PACM, Princeton University,\\
Princeton, New Jersey 08544, USA}
\author{Ping Sheng}
\address{Department of Physics and Institute of Nano Science and Technology,\\
Hong Kong University of Science and Technology,
Clear Water Bay, Kowloon, Hong Kong, China}
\maketitle
 
\begin{abstract}

We generalize the string method, originally designed for the study
of thermally activated rare events, to the calculation of quantum
tunneling rates. This generalization is based on the analogy
between quantum mechanics and statistical mechanics in 
the path-integral formalism. The quantum string method first locates,
in the space of imaginary-time trajectories, the minimal action path (MAP) 
between two minima of the imaginary-time action.
From the MAP, the saddle-point (``bounce'') action
associated with the exponential barrier penetration probability is
obtained and the pre-exponential factor (the ratio of determinants) for 
the tunneling rate evaluated using stochastic simulation.
The quantum string method is implemented to calculate the zero-temperature
escape rates for the metastable zero-voltage states 
in the current-biased Josephson tunnel junction model. 
In the regime close to the critical bias current,
direct comparison of the numerical and analytical results yields
good agreement. Our calculations indicate that for the nanojunctions
encountered in many experiments today, the (absolute) escape rates should
be measurable at bias current much below the critical current.

\end{abstract}

\section{Introduction}\label{intro}

One of the most important problems in statistical physics involves 
the rate of decay of a system rendered unstable by thermally activated 
barrier crossing \cite{kramers,landauer-swanson,langer}
and/or quantum barrier tunneling \cite{coleman,leggett,affleck}, 
and functional integrals represent a fundamental tool for studying these 
transition processes. 
However, numerical evaluation of the functional integrals 
has always been a challenge. Recently, the string method has been 
proposed for the numerical evaluation of thermally activated rare events
\cite{ren-prb,ren-jap,ERE,Ren,QRS}. 
This method first locates the most probable transition pathway
connecting two metastable states in configuration space. 
The transition rates can then be computed by numerically evaluating
the fluctuations around the most probable pathway.

The purpose of this paper is to generalize the string method to 
the study of quantum metastability caused by barrier tunneling. 
The theory of the decay rate through barrier tunneling has been formulated
using the imaginary-time functional integral techniques \cite{coleman}.
Essentially, the saddle point of the imaginary-time action is first
located and the rate of decay is then obtained by evaluating 
the relevant fluctuations. Within the functional integral formalism,
the computational task for a quantum field in $d$-dimensional 
space is equivalent to that for a classical field in $(d+1)$-dimensional 
space \cite{coleman}. Therefore, the quantum version of the string method 
can be numerically implemented as its original classical version 
for a higher dimensional system.

It should be noted that while in zero-dimensional tunneling problems 
(a quantum particle is regarded as a zero-dimensional quantum field,
as in the example presented below) the quantum string method might 
not offer any special advantage than, e.g., the well-known WKB method, 
in higher dimensional problems or in field theoretical formulations, 
the usual wave-mechanics approach becomes very difficult to implement.  
It is precisely in such problems that the present approach can offer 
an efficient numerical tool in finding the path of ``least action'', 
and on that basis calculate the relevant tunneling rate(s).

The paper is organized as follows. In Sec. \ref{string} we outline 
the string method for the numerical evaluation of thermal activation
rate and generalize it to the evaluation of quantum tunneling rate.
In Sec. \ref{jj} we apply the quantum string method to study 
the current-biased Josephson junction \cite{AH,fulton,voss,clarke}.
This physical model has long been used to demonstrate 
the quantum mechanical behavior of a single macroscopic degree of freedom 
(the phase difference across the tunnel junction) \cite{clarke}.
It has also played an important role in the study of 
macroscopic quantum tunneling \cite{MQT}. 
The escape rate of the junction from its zero-voltage state is 
numerically evaluated at zero temperature in the absence of dissipation. 
For bias currents less than but very close to the critical current, 
the tilted-washboard potential can be approximated by 
the quadratic-plus-cubic potential, for which the analytic form of
the quantum tunneling rate has been obtained 
\cite{caldeira-leggett,ueda-leggett}.
Our numerical results obtained for this ``solvable'' model 
show good agreement with the previous analytic results, and 
thus affirm the validity of the quantum string method. 
In Sec. \ref{disc} we conclude the paper with a few remarks
on the relationship of our results to quantum dissipation.

\section{Quantum string method}\label{string}

\subsection{String method for thermally activated rare events}
\label{background}

The string method was originally presented for the numerical study of 
thermal activation of metastable states \cite{ren-prb}. 
Consider a system governed by the energy potential $V({\bf q})$ 
in the overdamped regime, where ${\bf q}$ denotes 
the generalized coordinates $\{q_i\}$.
The minima of $V({\bf q})$ in the configuration space correspond to 
the metastable and stable states of the system.
Given ${\bf q}_A$ and ${\bf q}_C$ as the two minima of $V$,
the most probable fluctuation which can carry the system from ${\bf q}_A$ 
to ${\bf q}_C$ (or ${\bf q}_C$ to ${\bf q}_A$) corresponds to the lowest
intervening saddle point ${\bf q}_B$ between these two minima, with
the transition rate given by \cite{kramers,landauer-swanson,langer}
\begin{equation}\label{thermal-rate}
\Gamma_T({A\rightarrow C})=\displaystyle\frac{|\lambda_B|}{2\pi\eta}
\left[\displaystyle\frac{|\det H({\bf q}_B)|}{\det H({\bf q}_A)}\right]
^{-1/2}\exp\left\{-\displaystyle\frac{\Delta V}{k_BT}\right\},
\end{equation}
where $\eta$ is the frictional coefficient, $H({\bf q})$ denotes 
the Hessian of $V({\bf q})$, 
$\lambda_B$ is the negative eigenvalue of $H({\bf q}_B)$,
and $\Delta V=V({\bf q}_B)-V({\bf q}_A)$ is the energy barrier.
For the numerical evaluation of $\Gamma_T$, we define 
the minimal energy path (MEP) as a smooth curve ${\bf q}^\star(s)$ 
in configuration space. It connects ${\bf q}_A$ and ${\bf q}_C$ with
intrinsic parametrization such as arc length $s$, satisfying
\begin{equation}\label{mep-def}
\left(\nabla V\right)^\perp({\bf q}^\star)=0,
\end{equation}
where $\left(\nabla V\right)^\perp$ is the component of $\nabla V$ 
normal to the curve ${\bf q}^\star(s)$. Physically, the MEP is the most
probable pathway for thermally activated transitions between 
${\bf q}_A$ and ${\bf q}_C$. To numerically locate the MEP in
${\bf q}$ space, a string ${\bf q}(s,t)$ (a smooth curve with 
intrinsic parametrization by $s$) connecting ${\bf q}_A$ and ${\bf q}_C$ 
is evolved according to
\begin{equation}\label{string-evolution}
\left(\displaystyle\frac{d{\bf q}}{dt}\right)^\perp=
-\left(\nabla V\right)^\perp({\bf q}),
\end{equation}
where $(d{\bf q}/dt)^\perp$ denotes the velocity normal to 
the string \cite{ren-prb}. To enforce the desired parametrization, e.g.,
equal arc length, the string is reparametrized every a few time steps.
The stationary solution of Eq. (\ref{string-evolution}) satisfies 
Eq. (\ref{mep-def}) which defines the MEP. Once the MEP is determined, 
the intervening saddle point ${\bf q}_B$ is known. 
The negative eigenvalue $\lambda_B$ and the corresponding eigenvector
can be directly obtained from the MEP, and the ratio of the determinants 
in Eq. (\ref{thermal-rate}) can be numerically computed using 
a stochastic method \cite{ERE}.

The above scheme for the calculation of thermal activation rates
has many applications, e.g., the condensation of a supersaturated vapor, 
the realignment of a magnetic domain \cite{ren-prb,ren-jap}, 
and the decay of persistent current in one-dimensional superconductor 
\cite{LA,MH,QRS}.  All of these transition processes occur when the system 
undergoes a fluctuation that is large enough to initiate the transition.
Physically, the thermal activation rate $\Gamma_T$ becomes practically
unmeasurable as the temperature is sufficiently low ($k_BT\ll \Delta V$). 
However, even though thermodynamic fluctuations are suppressed at low
temperatures, a system can still be rendered unstable by quantum 
barrier tunneling \cite{coleman}. The simplest example is a particle 
that escapes a potential well: it penetrates a potential barrier and 
emerges at the escape point with zero kinetic energy, after which it 
propagates classically. In quantum field theory, a classical false vacuum 
is rendered unstable by bubbles of the true vacuum, realized through
tunneling.  Once these bubbles are sufficiently large, they become 
energetically favorable to grow.

\subsection{Rate of barrier tunneling}\label{rate}

The theory for the rate of barrier tunneling has been formulated 
\cite{coleman} and generalized \cite{leggett,affleck} using 
the imaginary-time functional integral techniques. Here we show that 
the string method can be generalized to be an efficient numerical tool 
(the quantum string method) for the calculation of tunneling rates.
In order to make concrete the formulation of the approach, below 
we consider a quantum particle that escapes a potential well 
through barrier tunneling. This zero-dimensional case
can be extended to a quantum field, where the classical false vacuum 
is rendered unstable by barrier tunneling \cite{coleman}.  
These results show that for a quantum field in $d$-dimensional space, 
the computational task may be reduced to the calculation of 
thermodynamic transition rates for a classical field in 
$(d+1)$-dimensional space.

\begin{figure}
\centerline{\psfig{figure=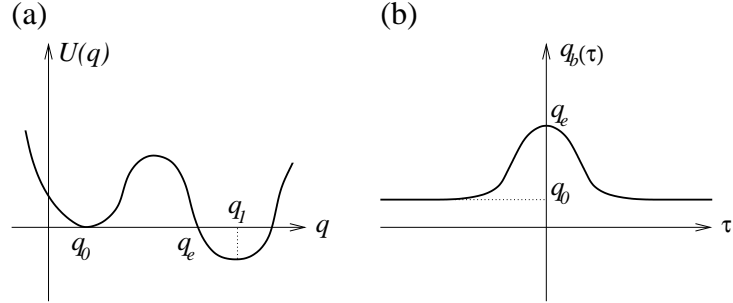,height=4cm}}
\bigskip
\caption{(a) The potential in one-dimensional space. 
The unstable ground state is centered at $q_0$. After penetrating 
the barrier, the particle emerges at the escape point $q_e$ and
propagates classically.
(b) The bounce solution $q_b(\tau)$ for the imaginary-time classical
equation of motion. In the inverted potential $-U(q)$, 
the particle begins at the top of the hill at $q_0$, turns around 
(i.e., bounces) at the classical turning point $q_e$, and returns to 
the top of the hill.}
\label{fig-potential-bounce}
\end{figure}

Consider a particle of mass $m$ moving in a one-dimensional potential 
$U(q)$ with two minima $q_0$ and $q_1$, one of which, $q_1$, 
is the absolute minimum (see Fig. \ref{fig-potential-bounce}a). 
Assume $U(q_0)=0$ and $q_1$ is to the right of 
$q_0$ ($q_1>q_0$). The normalized harmonic-oscillator ground state 
$\psi_g(q)$, centered at $q_0$, is 
$\psi_g(q)=\left({m\omega_0}/{\pi\hbar}\right)^{1/4}
\exp\left[-{m\omega_0}(q-q_0)^2/{2\hbar}\right]$, where 
$\omega_0=\sqrt{U''(q_0)/m}$ is the frequency locally defined at $q_0$.
The ground-state energy is $\hbar\omega_0/2$. These familiar results
can be derived from the imaginary-time propagator:
\begin{equation}\label{propagator0}
K(q_0,q_0;T)=\langle q_0|e^{-\hat{H}T/\hbar}|q_0\rangle
=\int [{\cal D}q(\tau)]e^{-S/\hbar},
\end{equation}
where $T$ is the imaginary time duration, 
$\hat{H}$ is the Hamiltonian of the system, 
\begin{equation}\label{action}
S[q(\tau)]=\int_{-T/2}^{T/2}d\tau\left\{\displaystyle\frac{m}{2}
\left(\displaystyle\frac{dq}{d\tau}\right)^2+U[q(\tau)]\right\}
\end{equation}
is the action, and $\int [{\cal D}q(\tau)]$ denotes the integration 
over functions $q(\tau)$ satisfying $q(-T/2)=q(T/2)=q_0$. 
Note that the imaginary-time (Euclidean) action $S[q(\tau)]$
can be obtained from the real-time (Minkowski) action
$${\cal A}[q(t)]=\int dt\left\{\displaystyle\frac{m}{2}\left[
\displaystyle\frac{dq(t)}{dt}\right]^2-U[q(t)]\right\}$$ 
through the formal substitution 
$t\rightarrow -i\tau$ and $-i{\cal A}[q(t)]\rightarrow S[q(\tau)]$. 
Thus the equation of motion in imaginary time would involve 
an inverted potential, i.e., $U(q)\rightarrow -U(q)$.
An expression for $K(q_0,q_0;T)$ in the limit of $T\rightarrow\infty$ 
gives both the energy and the wavefunction of the lowest-lying 
energy eigenstate. In the semiclassical (small $\hbar$) limit, 
the functional integral for $K(q_0,q_0;T)$ is dominated by 
the stationary points of the action, denoted by $\bar{q}(\tau)$, 
that satisfy the imaginary-time equation of motion 
$$\left[\displaystyle\frac{\delta S}{\delta q}\right]_{\bar{q}}=
-m\displaystyle\frac{d^2\bar{q}}{d\tau^2}+U'(\bar{q})=0,$$
with the boundary condition $\bar{q}(-T/2)=\bar{q}(T/2)=q_0$. 
There are two solutions: one is $\bar{q}(\tau)=q_0$, 
at which the Hessian of $S$, $-m\partial_\tau^2+U''(q_0)$, 
has positive eigenvalues only, and the other is the so-called 
bounce $q_b(\tau)$, at which the Hessian of $S$, 
$-m\partial_\tau^2+U''[q_b(\tau)]$, 
has a zero eigenvalue plus a negative eigenvalue. 
The zero eigenvalue comes from the time-translation symmetry and
the negative eigenvalue makes $q_b(\tau)$ a saddle point of $S[q(\tau)]$.
By following the bounce $q_b(\tau)$ in time, the quantum particle 
would initially stay at $q_0$ for a long time, on the order of $T$, 
then make a brief excursion to the escape point $q_e$ 
(separated from $q_0$ by a potential barrier, with $U(q_e)=U(q_0)$) 
in a time of order $1/\omega_0$, and finally return to $q_0$ and 
remain there for another duration of order $T$ 
(see Fig. \ref{fig-potential-bounce}b). 
This process is called a ``bounce''. Here $q_e$ is the coordinate point
at which the quantum tunneling particle leaves the barrier. 
Physically $q_b(\tau)$ characterizes a fluctuation 
large enough to accomplish the penetration. 
[For more details, see Appendix \ref{appendix-bounce}.]
Using the two stationary points of $S$ in the semiclassical approximation
with a proper analytic continuation \cite{coleman}, the propagator 
in Eq. (\ref{propagator0}) is obtained as
\begin{equation}\label{propagator}
|\psi_g(q_0)|^2e^{-E_gT/\hbar}=\left({m\omega_0}/{\pi\hbar}\right)^{1/2}
e^{-\omega_0T/2}e^{-i{\rm Im}E_gT/\hbar},
\end{equation}
which gives $|\psi_g(q_0)|^2=\left({m\omega_0}/{\pi\hbar}\right)^{1/2}$, 
the ground-state energy $\hbar\omega_0/2$, and an imaginary part of 
the energy ${\rm Im}E_g$. Physically, ${\rm Im}E_g$ ($<0$) is responsible 
for the decay of the metastable ``ground state'' centered at $q_0$, 
with the decay rate $\Gamma_Q=-2{\rm Im}E_g/\hbar$:
\begin{equation}\label{quantum-rate}
\Gamma_Q=\omega_0\sqrt{\displaystyle\frac{S_b}{2\pi\hbar}}\left[
\displaystyle\frac{U''(q_0)|\det'{\cal H}[q_b(\tau)]|}{\det{\cal H}(q_0)}
\right]^{-1/2}e^{-S_b/\hbar},
\end{equation}
where $\cal{H}$ denotes the Hessian of $S$:
${\cal H}[q(\tau)]=-m\partial_\tau^2+U''[q(\tau)]$, $S_b\equiv
S[q_b(\tau)]$ is the action associated with the bounce $q_b(\tau)$, 
and $\det'$ indicates that the zero eigenvalue is to be omitted in 
computing the determinant.

\subsection{Numerical implementation of quantum string method}
\label{numerical}

To numerically evaluate $\Gamma_Q$, we generalize the string method
to the quantum case. For formal similarity, we write the action 
$S[q(\tau)]$ in Eq. (\ref{action}) as $S({\bf q})$, where the vector
${\bf q}$ represents the coordinates in the $q(\tau)$-function space
(${\bf q}$ space). [Computationally, there are always a large but 
finite number of these coordinates.] We define the minimal action path 
(MAP) as a smooth curve ${\bf q}^\star(s)$ in ${\bf q}$ space. 
It connects the two minima of $S$, ${\bf q}_0$ and ${\bf q}_1$, with
intrinsic parametrization such as arc length $s$, satisfying
\begin{equation}\label{map-def}
\left(\nabla S\right)^\perp({\bf q}^\star)=0,
\end{equation}
where $\left(\nabla S\right)^\perp$ is the component of $\nabla S$ 
normal to the curve ${\bf q}^\star(s)$. Here ${\bf q}_0$ and ${\bf q}_1$ 
correspond to $\bar{q}(\tau)=q_0$ and $\bar{q}(\tau)=q_1$, respectively.
[A slightly different choice for ${\bf q}_1$ is also possible, 
corresponding to a $\bar{q}(\tau)$ profile with $\bar{q}(\tau)=q_1$
in most of the $\tau$-interval $[-T/2,T/2]$
and $\bar{q}(-T/2)=\bar{q}(T/2)=q_0$.]
The saddle point of $S$ is obtained as the point ${\bf q}_b$ which has 
the maximum value of $S$ along the MAP. 
This corresponds to the bounce $q_b(\tau)$.
To numerically locate the MAP in ${\bf q}$ space, a string 
${\bf q}(s,t)$ connecting ${\bf q}_0$ and ${\bf q}_1$ is evolved 
according to Eq. (\ref{string-evolution}) with $V$ replaced by $S$. 
The stationary solution is the MAP defined by Eq. (\ref{map-def}).

The ratio of determinants in Eq. (\ref{quantum-rate}) can be numerically
obtained as follows:\\
(1) From the MAP ${\bf q}^\star(s)$ parametrized by the arc length $s$,
the eigenvector ${\bf u}_b^{(1)}$ corresponding to the negative
eigenvalue $\lambda_b^{(1)}$ of the Hessian ${\cal H}({\bf q}_b)$ 
can be obtained by evaluating $d{\bf q}^\star(s)/ds$ at the saddle point
${\bf q}_b$, followed by a normalization. $\lambda_b^{(1)}$ is then 
computed from $\lambda_b^{(1)}
=[{\bf u}_b^{(1)}]^T{\cal H}({\bf q}_b){\bf u}_b^{(1)}$. \\
(2) The eigenvector ${\bf u}_b^{(2)}$ corresponding to the zero
eigenvalue $\lambda_b^{(2)}$ of the Hessian ${\cal H}({\bf q}_b)$ 
can be obtained by evaluating $\partial_\tau q_b(\tau)$ from ${\bf q}_b$, 
followed by a normalization.\\
(3) The Hessian ${\cal H}({\bf q}_b)$ is modified to give a positive
definite matrix $\tilde{\cal H}({\bf q}_b)$:
\begin{equation}\label{modifyH_b}
\tilde{\cal H}({\bf q}_b)={\cal H}({\bf q}_b)+2|\lambda_b^{(1)}|
[{\bf u}_b^{(1)}][{\bf u}_b^{(1)}]^T+
[{\bf u}_b^{(2)}][{\bf u}_b^{(2)}]^T,
\end{equation}
whose determinant $\det\tilde{\cal H}({\bf q}_b)$ equals 
$|\det'{\cal H}({\bf q}_b)|$.\\
(4) In order to compute the ratio of determinants 
${\det\tilde{\cal H}({\bf q}_b)}/{\det{\cal H}({\bf q}_0)}$, 
a harmonic potential parametrized by $\alpha$ ($0\le\alpha\le 1$)
is constructed as
\begin{equation}\label{potential-U}
{\cal U}^\alpha({\bf q})=\displaystyle\frac{1}{2}
{\bf q}^T[(1-\alpha)\tilde{\cal H}({\bf q}_b)
+\alpha {\cal H}({\bf q}_0)]{\bf q}
\end{equation}
in the $N$-dimensional ${\bf q}$ space.
It can be shown \cite{ERE} that 
\begin{equation}\label{determinant-ratio}
\displaystyle\frac{\det\tilde{\cal H}({\bf q}_b)}
{\det{\cal H}({\bf q}_0)}=\exp\left[2\int_0^1Q(\alpha) d\alpha \right],
\end{equation}
where $Q(\alpha)$ is the expectation value 
\begin{equation}\label{expectation}
Q(\alpha)=\displaystyle\frac{1}{Z(\alpha)}\int d{\bf q}
\left\{\displaystyle\frac{1}{2}{\bf q}^T\left[
\tilde{\cal H}({\bf q}_b)-{\cal H}({\bf q}_0)\right]{\bf q}\right\}\exp
\left[-{\cal U}^\alpha({\bf q})\right],
\end{equation}
in which $Z(\alpha)$ is the partition function
\begin{equation}\label{partition-Z}
Z(\alpha)=\int d{\bf q}\exp\left[-{\cal U}^\alpha({\bf q})\right].
\end{equation}
A stochastic process can be generated to measure $Q(\alpha)$
according to Eq. (\ref{expectation}) \cite{ERE}.

\section{Current-biased josephson tunnel junction}\label{jj}

\subsection{Model formulation}\label{model}

\begin{figure}
\centerline{\psfig{figure=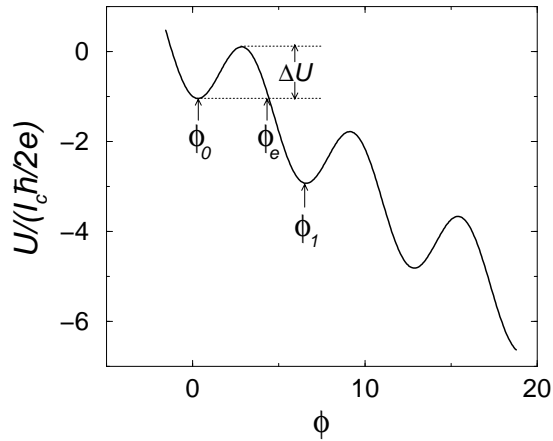,height=6cm}}
\bigskip
\caption{The tilted washboard potential $U(\phi)$ for $I/I_c=0.3$.
Here the unstable ground state is centered at $\phi_0$, $\phi_e$ 
is the escape point, and $\phi_1=\phi_0+2\pi$ is the next minimum.}
\label{fig-washboard}
\end{figure}

Superconducting devices based on the Josephson effect have been widely 
used to investigate macroscopic quantum tunneling \cite{MQT}.
We consider the resistively and capacitively shunted junction 
\cite{AH,voss,clarke} for which the classical equation of motion is
\begin{equation}\label{RCSJ}
C\left(\displaystyle\frac{\hbar}{2e}\right)^2
\displaystyle\frac{d^2\phi}{dt^2}
+\displaystyle\frac{1}{R}\left(\displaystyle\frac{\hbar}{2e}\right)^2
\displaystyle\frac{d\phi}{dt}+\displaystyle\frac{\partial }{\partial\phi}
\left(-\displaystyle\frac{I_c\hbar}{2e}\cos\phi-
\displaystyle\frac{I\hbar}{2e}\phi\right)=
\displaystyle\frac{\hbar}{2e}I_N(t),
\end{equation}
where $\phi$ is the difference in the phases of the order parameters
on two sides of the junction, $C$ is the capacitance of the junction,
$R$ is the resistance of the junction, $I_c$ is the Josephson 
critical current, $I$ is the bias current, and $I_N(t)$ is
the fluctuating noise current generated by $R$.
Equation (\ref{RCSJ}) is the same as that of a particle of mass
$C(\hbar/2e)^2$ moving along the $\phi$ axis in an effective potential
(the so-called ``tilted washboard'' potential, see Fig. \ref{fig-washboard})
$$U(\phi)=-\displaystyle\frac{I_c\hbar}{2e}
\left[\cos\phi+\displaystyle\frac{I}{I_c}\phi\right].$$
According to this mechanical analog, for $I<I_c$ the zero-voltage
state is given by $\phi_0=\arcsin(I/I_c)$, which is a minimum of
$U(\phi)$. In the classical limit, the escape from 
the zero-voltage state is induced by the noise current which activates
the system over the potential barrier \cite{AH}. 
At sufficiently low temperatures, although thermodynamic fluctuations 
are suppressed, the junction can still escape from the zero-voltage state 
through quantum barrier tunneling \cite{voss,clarke}.

We consider the zero-temperature behavior of the junction
in the low-damping limit 
($(2eI_c/\hbar C)^{1/2}[1-(I/I_c)^2]^{1/4}RC\gg 1$). 
The state of the junction is represented by the wave function
$\Psi(\phi,t)$, governed by the Schr\"odinger equation
$i\hbar\partial\Psi/\partial t=\hat{H}\Psi$ in which the Hamiltonian
is of the form
$$\hat{H}=-\displaystyle\frac{(2e)^2}{2C}\displaystyle\frac
{\partial^2}{\partial\phi^2}+U(\phi).$$
We introduce three parameters: the dimensionless bias current $x=I/I_c$, 
the plasma frequency $\omega_p=\sqrt{2eI_c/\hbar C}$, and 
$I_c/2e\omega_p=\sqrt{E_J/E_C}$ where $E_J=I_c\hbar/2e$ is 
the Josephson coupling energy and $E_C=(2e)^2/C$ is the charging energy.
The imaginary-time action 
\begin{equation}\label{jj-action1}
S[\phi(\tau)]=\int d\tau\left[\displaystyle\frac{1}{2}
C\left(\displaystyle\frac{\hbar}{2e}\right)^2
\left(\displaystyle\frac{d\phi}{d\tau}\right)^2+U(\phi)\right],
\end{equation}
can be written as $S[\phi(\tau)]=\hbar\sqrt{E_J/E_C}
\bar{S}[\phi(\bar{\tau})]$, where $\bar{\tau}=\omega_p\tau$ is 
a dimensionless time variable and $\bar{S}$ is the dimensionless action
\begin{equation}\label{jj-action2}
\bar{S}[\phi(\bar{\tau})]=\int d\bar{\tau}\left[\displaystyle\frac{1}{2}
\left(\displaystyle\frac{d\phi}{d\bar{\tau}}\right)^2+
\left(-\cos\phi-x\phi\right)\right].
\end{equation}

\subsection{Numerical results}\label{data}

Based on the actions (\ref{jj-action1}) and (\ref{jj-action2}), 
the tunneling rate is given by
\begin{equation}\label{jj-rate}
\Gamma_Q=\omega_p\sqrt{\cos\phi_0}
\left(\displaystyle\frac{E_J}{E_C}\right)^{1/4}
\sqrt{\displaystyle\frac{\bar{S}_b}{2\pi}}
\left[\displaystyle\frac{\cos\phi_0|\det'{\cal H}[\phi_b(\bar{\tau})]|}
{\det{\cal H}(\phi_0)}\right]^{-1/2}\exp
\left(-\sqrt{\displaystyle\frac{E_J}{E_C}}\bar{S}_b\right),
\end{equation}
according to the general expression (\ref{quantum-rate}).
Here $\omega_p\sqrt{\cos\phi_0}$ is the frequency locally defined at
$\phi_0$,
$\bar{S}_b\equiv\bar{S}[\phi_b(\bar{\tau})]$ is 
the dimensionless action of the bounce $\phi_b(\bar{\tau})$ 
which satisfies $\phi_b(\pm\infty)=\phi_0$, 
and ${\cal H}$ is the Hessian of $\bar{S}[\phi(\bar{\tau})]$, given by
${\cal H}[\phi(\bar{\tau})]=-\partial^2/\partial\bar{\tau}^2
+\cos[\phi(\bar{\tau})]$.
Numerical calculations based on the action in Eq. (\ref{jj-action2}) 
have been carried out to evaluate the bounce $\phi_b(\bar{\tau})$,
the bounce action $\bar{S}_b$, and the ratio of determinants in the general 
expression (\ref{jj-rate}). Note that these dimensionless properties
are uniquely determined by the parameter $x$. Once they are evaluated,
the tunneling rate $\Gamma_Q$ can be readily obtained using 
the other two parameters $\omega_p$ and $\sqrt{E_J/E_C}$.
These parameters should be easily measurable experimentally, since
they are directly determined from the capacitance of the junction, 
the critical current, and the bias current.

Numerical calculations have been carried out according to the following
procedure.

\begin{figure}
\centerline{\psfig{figure=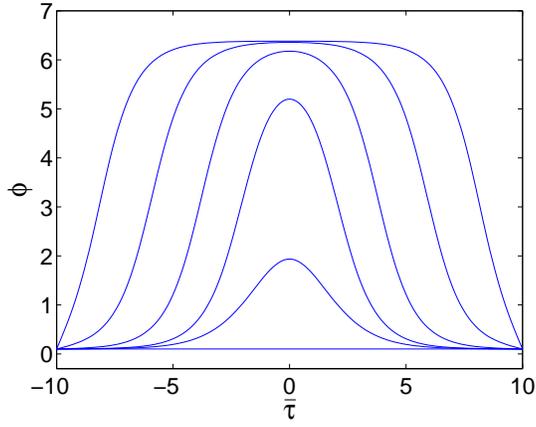,height=6cm}}
\bigskip
\caption{A sequence of the $\phi(\bar{\tau})$ profiles along the MAP for 
$x=0.1$. From bottom to top, the first curve denotes 
$\phi(\bar{\tau})=\phi_0$ which is a local minimum of 
$\bar{S}[\phi(\bar{\tau})]$, the third curve denotes the bounce 
$\phi_b(\bar{\tau})$, and the last curve denotes another local minimum of 
$\bar{S}[\phi(\bar{\tau})]$ with $\phi(0)=\phi_1$ and 
$\phi(-\bar{T}/2)]=\phi(\bar{T}/2)=\phi_0$.}
\label{fig-mep}
\end{figure}

\begin{figure}
\centerline{\psfig{figure=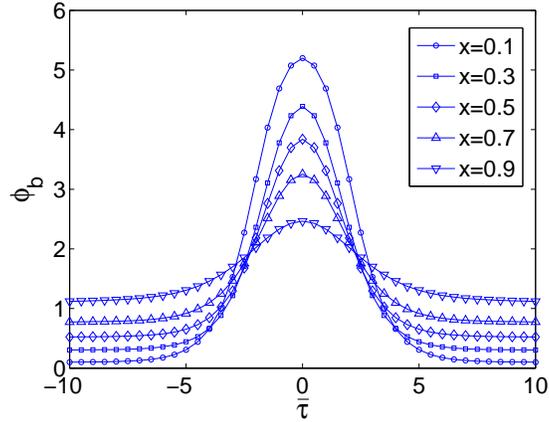,height=6cm}}
\bigskip
\caption{The bounce $\phi_b(\bar{\tau})$, plotted for a few selected values
of $x$.}
\label{fig-bounce}
\end{figure}

\begin{figure}
\centerline{\psfig{figure=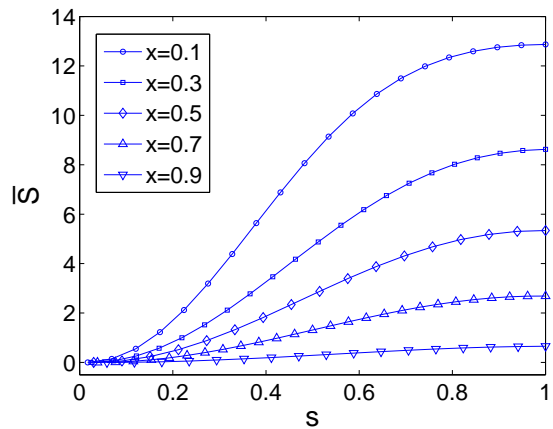,height=6cm}}
\bigskip
\caption{The dimensionless action along a segment of the MAP
starting from $\phi(\bar{\tau})=\phi_0$ and ending at $\phi_b(\bar{\tau})$.
Here $\bar{S}$ is plotted as a function of the arc length $s$ in 
the $\phi(\bar{\tau})$-function space for a few selected values of $x$.
The profile of $\phi(\bar{\tau})=\phi_0$ is taken as the reference point 
$s=0$ at which $\bar{S}$ has been set to be zero by a constant shift of 
the potential.}
\label{fig-map-action}
\end{figure}

\begin{figure}
\centerline{\psfig{figure=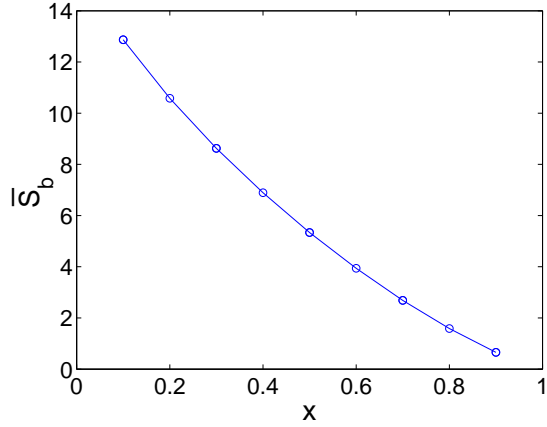,height=6cm}}
\bigskip
\caption{The bounce action $\bar{S}_b$ plotted as a function of $x$. 
}
\label{fig-bounce-action}
\end{figure}

(i) We first locate the MAP in the $\phi(\bar{\tau})$-function space.
In the calculation, $\phi(\bar{\tau})$ is represented by a column vector 
${\mbox{\boldmath$\phi$}}$ of $N=200$ entries, with the $\bar{\tau}$-interval 
$[-\bar{T}/2,\;\bar{T}/2]$ discretized by a uniform mesh of $N$ points.
We use $\bar{T}=20$, large enough for the computation of zero-temperature
properties. [Here $\bar{T}=\hbar\omega_p/k_B{\cal T}$, where $k_B$ is 
the Boltzmann constant and ${\cal T}$ the temperature, and $\bar{T}\gg 1$
means $\hbar\omega_p\gg k_B{\cal T}$.]
The string ${\mbox{\boldmath$\phi$}}(s)$ connecting 
${\mbox{\boldmath$\phi$}}(0)={\mbox{\boldmath$\phi$}}_0$
and ${\mbox{\boldmath$\phi$}}(1)={\mbox{\boldmath$\phi$}}_1$
is discretized by $M=200$ points in the ${\mbox{\boldmath$\phi$}}$ space.
As to the two fixed ends of the string, ${\mbox{\boldmath$\phi$}}_0$ 
corresponds to $\phi(\bar{\tau})=\phi_0$ and ${\mbox{\boldmath$\phi$}}_1$ 
to a $\phi(\bar{\tau})$ profile with $\phi(\bar{\tau})=\phi_1$ in most
of the $\bar{\tau}$-interval and $\phi(-\bar{T}/2)=\phi(\bar{T}/2)=\phi_0$.
Here $\phi_0=\arcsin x$ and $\phi_1=\phi_0+2\pi$ are two neighboring
minima of $-\left(\cos\phi+x\phi\right)$ (see Fig. \ref{fig-washboard}).
Note that ${\mbox{\boldmath$\phi$}}_1$ is obtained as a local minimum
of $\bar{S}$ in Eq. (\ref{jj-action2}).
The string evolution is generated by
$$\left({d{\mbox{\boldmath$\phi$}}}/{dt}\right)^\perp=
-\left(\nabla\bar{S}\right)^\perp({\mbox{\boldmath$\phi$}}),$$ 
with the initial string taken from a linear interpolation 
between ${\mbox{\boldmath$\phi$}}_0$ and ${\mbox{\boldmath$\phi$}}_1$.
The MAP ${\mbox{\boldmath$\phi$}}^\star(s)$ is reached by 
the evolving string ${\mbox{\boldmath$\phi$}}(s,t)$
as its stationary solution defined by 
$\left(\nabla\bar{S}\right)^\perp({\mbox{\boldmath$\phi$}}^\star)=0$,
with ${\mbox{\boldmath$\phi$}}^\star(0)={\mbox{\boldmath$\phi$}}_0$
and ${\mbox{\boldmath$\phi$}}^\star(1)={\mbox{\boldmath$\phi$}}_1$.
The bounce $\phi_b(\bar{\tau})$ is obtained from the vector
${\mbox{\boldmath$\phi$}}_b$ which yields the maximum value of $\bar{S}$ 
along the MAP.
In Fig. \ref{fig-mep} a sequence of the $\phi(\bar{\tau})$ profiles
along the MAP is shown for $x=0.1$, and
in Fig. \ref{fig-bounce} the bounce profile $\phi_b(\bar{\tau})$
is shown for a few selected values of $x$.
In Fig. \ref{fig-map-action} the variation of the action $\bar{S}$ along 
the MAP is shown for a few selected values of $x$, and 
in Fig. \ref{fig-bounce-action} the bounce action $\bar{S}_b$ is plotted as 
a function of $x$.

\begin{figure}
\centerline{\psfig{figure=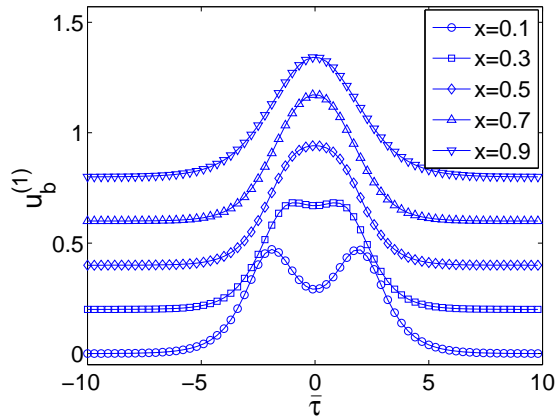,height=6cm}}
\bigskip
\caption{The eigenfunction $u_b^{(1)}(\bar{\tau})$ of the Hessian 
${\cal H}[\phi_b(\bar{\tau})]$ with the negative eigenvalue, 
plotted for a few selected values of $x$.
The curves are displaced vertically for clarity, 
whereas the original ones all start and end at $0$.
Each eigenfunction represents the unstable direction at the saddle point
along a particular MAP in the $\phi(\bar{\tau})$-function space.
It is noted that $u_b^{(1)}(\bar{\tau})$ obtained for $x=0.1$ 
is qualitatively different from that for $x=0.9$. 
For $x$ close to $0$, the bounce $\phi_b(\bar{\tau})$ has the height
$\phi_e$ (the escape point) close to the next (lower) minimum $\phi_1$,
and the growth of the $\phi(\bar{\tau})$ bubble along the MAP is 
characterized by the movement of the two (left and right) domain walls.
Consequently, the unstable direction $u_b^{(1)}(\bar{\tau})$ shows two peaks.
On the other hand, for $x>0.3$, the bounce $\phi_b(\bar{\tau})$ 
has the height $\phi_e$ far from the next minimum $\phi_1$, 
and the growth of the $\phi(\bar{\tau})$
bubble along the MAP is characterized by overall dilation. Consequently,
the unstable direction $u_b^{(1)}(\bar{\tau})$ displays one peak only.}
\label{fig-negative}
\end{figure}

(ii)  We then modify the Hessian ${\cal H}[\phi_b(\bar{\tau})]$, 
represented by the $N\times N$ matrix
${\cal H}({\mbox{\boldmath$\phi$}}_b)$, to give a positive
definite matrix $\tilde{\cal H}({\mbox{\boldmath$\phi$}}_b)$, given by
$$\tilde{\cal H}({\mbox{\boldmath$\phi$}}_b)=
{\cal H}({\mbox{\boldmath$\phi$}}_b)+2|\lambda_b^{(1)}|
[{\bf u}_b^{(1)}][{\bf u}_b^{(1)}]^T+
\cos\phi_0[{\bf u}_b^{(2)}][{\bf u}_b^{(2)}]^T.$$
Here ${\bf u}_b^{(1)}$ is the eigenvector corresponding to the negative
eigenvalue $\lambda_b^{(1)}$ of the Hessian 
${\cal H}({\mbox{\boldmath$\phi$}}_b)$, 
${\bf u}_b^{(2)}$ is the eigenvector corresponding to the zero
eigenvalue $\lambda_b^{(2)}$ of the same Hessian, and 
$\det\tilde{\cal H}({\mbox{\boldmath$\phi$}}_b)=
\cos\phi_0|\det'{\cal H}({\mbox{\boldmath$\phi$}}_b)|$.
Note that ${\bf u}_b^{(1)}$, $\lambda_b^{(1)}$, and ${\bf u}_b^{(2)}$
can be readily obtained once the MAP is determined, as outlined
in Sec. \ref{numerical}. In Fig. \ref{fig-negative}
the unstable direction ${\bf u}_b^{(1)}$ along the MAP 
at the saddle point is shown for a few selected values of $x$.

\begin{figure}
\centerline{\psfig{figure=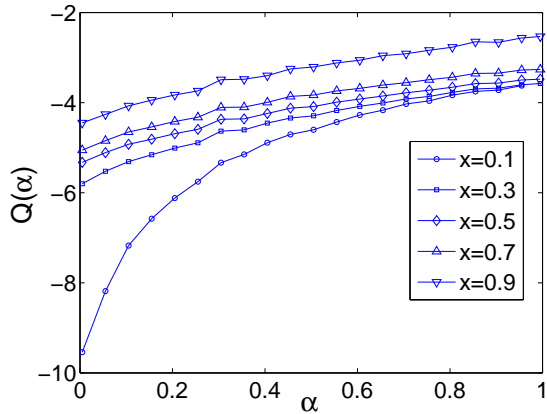,height=6cm}}
\bigskip
\caption{Stochastically measured $Q(\alpha)$,
plotted for a few selected values of $x$.}
\label{fig-Q}
\end{figure}

\begin{figure}
\centerline{\psfig{figure=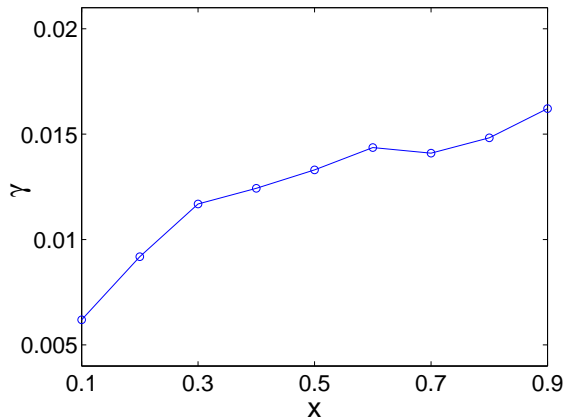,height=6cm}}
\bigskip
\caption{The determinant ratio $\gamma$ in Eq. (\ref{ratio}) 
plotted as a function of $x$. 
}
\label{fig-ratio}
\end{figure}

\begin{figure}
\centerline{\psfig{figure=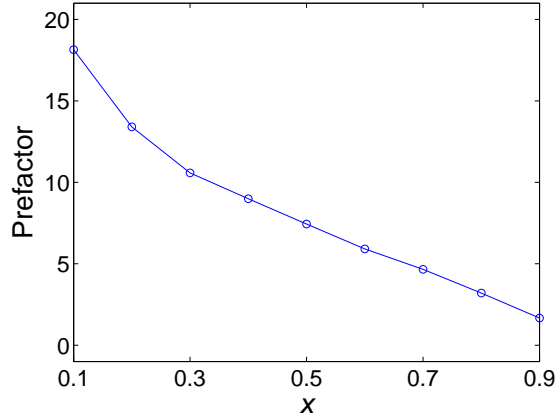,height=6cm}}
\bigskip
\caption{The dimensionless prefactor 
$\sqrt{\cos\phi_0}\sqrt{\displaystyle\frac{\bar{S}_b}{2\pi}}
\left[\displaystyle\frac{\cos\phi_0|\det'{\cal H}[\phi_b(\bar{\tau})]|}
{\det{\cal H}(\phi_0)}\right]^{-1/2}$
in Eq. (\ref{jj-rate}), plotted as a function of $x$.
Note the large variation of a factor $>10$ as a function of $x$.
}
\label{fig-prefactor}
\end{figure}

(iii)  We calculate the ratio of determinants 
\begin{equation}\label{ratio}
\gamma=\displaystyle\frac{\cos\phi_0|\det'{\cal H}[\phi_b(\bar{\tau})]|}
{\det{\cal H}(\phi_0)}=
\displaystyle\frac{\cos\phi_0|\det'{\cal H}({\mbox{\boldmath$\phi$}}_b)|}
{\det{\cal H}({\mbox{\boldmath$\phi$}}_0)},
\end{equation}
where ${\cal H}({\mbox{\boldmath$\phi$}}_0)$ is the $N\times N$ matrix
representation of ${\cal H}(\phi_0)$.
This is done by evaluating $\det\tilde{\cal H}({\mbox{\boldmath$\phi$}}_b)
/\det{\cal H}({\mbox{\boldmath$\phi$}}_0)$ according to 
the stochastic method outlined in Sec. \ref{numerical}.
Numerical results of this part are shown in Figs. \ref{fig-Q} and
\ref{fig-ratio}, for the stochastically measured $Q(\alpha)$ in 
Eqs. (\ref{determinant-ratio}) and (\ref{expectation})
and the ratio of determinants $\gamma$ in Eq. (\ref{ratio}). 
From $\bar{S}_b$ and $\gamma$, the dimensionless prefactor
$\sqrt{\cos\phi_0}\sqrt{\displaystyle\frac{\bar{S}_b}{2\pi}}
\left[\displaystyle\frac{\cos\phi_0|\det'{\cal H}[\phi_b(\bar{\tau})]|}
{\det{\cal H}(\phi_0)}\right]^{-1/2}$ in Eq. (\ref{jj-rate})
is readily obtained and plotted in Fig. \ref{fig-prefactor}.

Using the numerical results for $\bar{S}_b$ and $\gamma$, 
the mean escape rate out of the zero-voltage state can be readily obtained
from Eq. (\ref{jj-rate}), once the values of $\omega_p$ and $\sqrt{E_J/E_C}$
are given. From $I_c=9.489\;\mu{\rm A}$ and $C=6.35\;{\rm pF}$ reported 
in an early experiment \cite{clarke}, 
we have $\omega_p=67.4\;{\rm GHz}$ and $\sqrt{E_J/E_C}=440$.
The largeness of $\sqrt{E_J/E_C}$ implies that quantum tunneling becomes
observable only if $x\rightarrow 1$ and $\bar{S}_b\rightarrow 0$. 
For the experiment reported in Ref. \cite{clarke}, $x\approx 0.99$ 
at which $\bar{S}_b\approx 0.037$,
$e^{-\sqrt{E_J/E_C}\bar{S}_b}\sim 10^{-7}$, and the escape rate
is approximately $2.7\times 10^4\;{\rm sec}^{-1}$. 
In a recent experiment on quantum superposition of macroscopic 
persistent-current states, a superconducting loop is constructed 
with three Josephson junctions \cite{superposition}. We find that
the junction parameters in that experiment allow quantum tunneling
to be observable in a range of $x$ much wider than that in Ref. \cite{clarke}.
From $I_c=570\;{\rm nA}$ and $C=2.6\;{\rm fF}$ \cite{superposition},
we have $\omega_p=816\;{\rm GHz}$ and $\sqrt{E_J/E_C}=2.18$.
The smallness of $\sqrt{E_J/E_C}$ then allows $\bar{S}_b$ and hence $x$
to vary in a wide range. For $x$ decreasing from $0.8$ to $0.2$,
$\bar{S}_b$ roughly increases from $2$ to $10$, and consequently, 
$e^{-\sqrt{E_J/E_C}\bar{S}_b}$ changes from
$\sim 10^{-2}$ to $\sim 10^{-10}$.
Using Eq. (\ref{jj-rate}) with the numerical results for
$\bar{S}_b$ and $\gamma$, we obtain the escape rates 
$\Gamma_Q=1.2\times 10^{11}\;{\rm sec}^{-1}$,
$7.4\times 10^{7}\;{\rm sec}^{-1}$, and 
$1.5\times 10^{3}\;{\rm sec}^{-1}$, 
for $x=0.8$, $0.5$, and $0.2$, respectively.
Note that typically, the measured escape rates are in the range 
from $\sim 10^1$ to $\sim 10^6\;{\rm sec}^{-1}$.
Therefore, our numerical results indicate that the absolute 
escape rates for today's nanojunctions should be measurable 
at bias current much below the critical current. This is because
the junction capacitance has been significantly reduced and thus
the action scale $\sqrt{E_J/E_C}$ can be made small enough to allow 
a relatively large dimensionless action $\bar{S}_b$ in the exponential 
factor $e^{-\sqrt{E_J/E_C}\bar{S}_b}$ \cite{charging-energy}. 
In this regard a numerical scheme as presented in this paper is essential 
to the evaluation of the absolute escape rates.

\subsection{Quadratic-plus-cubic potential}\label{quadr-cubic}

For bias currents less than but very close to $I_c$ ($x$ close to $1$),
the potential barrier $\Delta U$ to be penetrated is low, 
the distance between the minimum $\phi_0$ and the escape point $\phi_e$
is small, and hence the potential $U(\phi)$ in the classically forbidden
region can be approximated by the quadratic-plus-cubic potential. 
That is, the dimensionless potential $-\cos\phi-x\phi$ in the action
(\ref{jj-action2}) can be expressed in a Taylor expansion form
\begin{equation}\label{approx-u}
u(\phi)=-\cos\phi-x\phi\approx (-\cos\phi_0-x\phi_0)
+\displaystyle\frac{1}{2}\cos\phi_0(\phi-\phi_0)^2
\left(1-\displaystyle\frac{\phi-\phi_0}{\phi_e-\phi_0}\right),
\end{equation}
where $\phi_e-\phi_0=3\cot\phi_0$ approaches zero as
$x\rightarrow 1$ and $\phi_0\rightarrow\pi/2$.
From Eq. (\ref{approx-u}), the dimensionless potential barrier can be 
easily found to be $\Delta u=2\cos\phi_0(\phi_e-\phi_0)^2/27$
and the dimensionless bounce action to be
$\bar{S}_b=8\sqrt{\cos\phi_0}(\phi_e-\phi_0)^2/15$.
In addition, the Hessian of $\bar{S}[\phi(\bar{\tau})]$ becomes
${\cal H}[\phi(\bar{\tau})]=-\partial^2/\partial\bar{\tau}^2
+\cos\phi_0[1-3(\phi-\phi_0)/(\phi_e-\phi_0)]$, for which 
the analytic result
$$\displaystyle\frac{\cos\phi_0|\det'{\cal H}[\phi_b(\bar{\tau})]|}
{\det{\cal H}(\phi_0)}=\displaystyle\frac{1}{60}$$
has been obtained for the determinant ratio $\gamma$ in Eq. (\ref{ratio}) 
\cite{caldeira-leggett,ueda-leggett}.
These exact results for the quadratic-plus-cubic potential allow 
an analytic form for the tunneling rate $\Gamma_Q$ in 
Eq. (\ref{quantum-rate}).
Numerical results for $x=0.9$ in Figs. \ref{fig-bounce-action} and 
\ref{fig-ratio} show that
$\displaystyle\frac{\bar{S}_b}{\sqrt{\cos\phi_0}(\phi_e-\phi_0)^2}=0.469$,
approaching $8/15$, and the determinant ratio $\gamma=0.0162$, approaching
$1/60$, though $\phi_e-\phi_0=1.45$ is still large.

In order to demonstrate the validity and precision of the quantum 
string method, numerical calculations have been carried out to reproduce 
the bounce action and determinant ratio for the quadratic-plus-cubic 
potential, an important model potential for the study of 
quantum metastability \cite{ueda-leggett}. 
For simplicity, we work with the scaled action functional
\begin{equation}\label{action23}
S[q(\tau)]=\int d\tau\left[\displaystyle\frac{1}{2}
\left(\displaystyle\frac{dq}{d\tau}\right)^2+\displaystyle\frac{1}{2}
q^2\left(1-q\right)\right].
\end{equation}
The potential $q^2\left(1-q\right)/2$ in action (\ref{action23}) 
has $q_0=0$ as the metastable minimum and $q_e=1$ as the escape point.
For computational purpose, this potential is distorted in the region of
$q\gg q_e$ to generate another (lower) minimum at $q_1$ ($\gg q_e$).
Numerically, the two potential minima $q_0$ and $q_1$ are used to
fix the ends of the evolving string in the $q(\tau)$-function space. 
The stationary solution for 
the string evolution equation is the MAP from which the saddle point
of $S[q(\tau)]$, i.e., the bounce $q_b(\tau)$, can be obtained. 
Since the potential profile is untouched in the classically forbidden 
region ($q_0\le q\le q_e$), the bounce so obtained is not affected by 
the potential distortion far away. The bounce action $S_b\equiv
S[q_b(\tau)]$ is obtained to be $0.5337$, very close to the exact result
$8/15$. The determinant ratio
\begin{equation}\label{ratio23}
\displaystyle\frac{|\det'{\cal H}[q_b(\tau)]|}{\det{\cal H}(q_0)}
=\displaystyle\frac{|\det'\left[-\partial^2/\partial\tau^2+
1-3q_b(\tau)\right]|}{\det\left(-\partial^2/\partial\tau^2+1\right)}
\left(=\displaystyle\frac{1}{60}\right)
\end{equation}
is obtained to be $0.0142$, close to $1/60$.
These results are obtained for $N=200$, $M=200$, and the total 
imaginary time duration $T=20$. Better agreement with the exact results
can certainly be achieved by using longer imaginary time duration,
vector space of higher dimensionality, and finer resolution in
discretizing the string.

\section{Concluding remarks}\label{disc}

Quantum tunneling in macroscopic systems is intimately related to 
the important issue of quantum dissipation, which arises from 
the coupling to environmental variables. 
This coupling can modify the tunneling itself, as many prior works 
have shown \cite{leggett,caldeira-leggett,freidkin,weiss}. 
However, it should be noted that regardless of the particulars 
in the quantum dissipation model, the net result is to decrease 
the escape rate. 
Hence the rate calculated for nondissipative quantum tunneling may be
regarded as an upper bound to the rate(s) with nonzero dissipation.
                                                                                
In this regard it should also be noted that as a tool for the numerical
evaluation of tunneling rate in the path integral formalism, 
the quantum string method is directly generalizable to field 
theoretic problems, requiring only additional computational resources.

\section*{Acknowledgments}
T. Q. was supported by the Hong Kong RGC under grant 602805.
J. S. and W. E. were supported by DOE under grant DE-FG02-03ER25587.
P. S. was supported by the Hong Kong RGC under grants HKUST6073/02P 
and CA04/05.SC02.

\appendix
\section{The bounce}\label{appendix-bounce}

The bounce $q_b(\tau)$ is a solution of the imaginary-time classical 
equation of motion: 
$$m\displaystyle\frac{d^2\bar{q}}{d\tau^2}=U'(\bar{q}),$$
subject to the boundary condition 
$\bar{q}(-T/2)=\bar{q}(T/2)=q_0$ for $T\rightarrow\infty$. 
The qualitative behavior of $q_b(\tau)$ is suggested by the analogy
with the equation of motion for a particle of mass $m$ in the inverted
potential $-U(q)$, in which $q_0$ now corresponds to the top of the hill 
and $q_e$ to the classical turning point. The particle 
would spend most of its time at $q_0$ (due to zero velocity), 
but, in the course of an arbitrarily long interval of time, it would 
make a brief excursion to the point $q_e$ and then return to $q_0$ 
(see Fig. \ref{fig-potential-bounce}b). Note that
$$\displaystyle\frac{m}{2}\left(\displaystyle\frac
{d\bar{q}}{d\tau}\right)^2-U[\bar{q}(\tau)]=-U(q_0)=0$$
is a constant of motion for $q_b(\tau)$. 
This means $dq_b/d\tau$ vanishes at $q_0$ and $q_e$.

The bounce $q_b(\tau)$ shown in Fig. \ref{fig-potential-bounce}b is centered
at $\tau_c=0$ along the $\tau$ axis. Because of the time-translation
invariance of the action, the bounce solutions are also given 
by $q_b(\tau-\tau_c)$, where $\tau_c$ is an arbitrary center of the bounce.
This symmetry property leads to a zero eigenvalue for the Hessian of $S$
at $q_b(\tau)$, $-m\partial_\tau^2+U''[q_b(\tau)]$. 
The corresponding eigenfunction is given by
$$u_b^{(2)}(\tau)=\sqrt{\displaystyle\frac{m}{S_b}}
\displaystyle\frac{dq_b}{d\tau},$$
where $\sqrt{{m}/{S_b}}$ is the normalization factor derived from
the action of the bounce,
$$S_b=\int_{-T/2}^{T/2}d\tau\left\{\displaystyle\frac{m}{2}
\left(\displaystyle\frac{dq_b}{d\tau}\right)^2+U[q_b(\tau)]\right\},$$
and the constant of motion 
$$\displaystyle\frac{m}{2}\left(\displaystyle\frac{dq_b}{d\tau}\right)^2
-U[q_b(\tau)]=0.$$
Note that $dq_b/d\tau$ has a zero at the center of the bounce.
Therefore, $u_b^{(2)}(\tau)$ has a node and can not be the eigenfunction
with the lowest eigenvalue: there must be a nodeless eigenfunction, 
$u_b^{(1)}(\tau)$, with a negative eigenvalue. This implies that 
the bounce is not a minimum of the action but a saddle point.
The negative eigenvalue requires a proper analytical continuation
in evaluating the functional integral in (\ref{propagator0}).
This leads to a complex energy in (\ref{propagator}).

Using the constant of motion, the bounce action can be reduced to 
the form
$$S_b=\int_{-T/2}^{T/2}m\left(\displaystyle\frac{dq_b}{d\tau}\right)^2
d\tau=2\int_{q_0}^{q_e}\sqrt{2mU(q)}dq.$$
It is seen that $e^{-S_b/2\hbar}$ is the familiar WKB exponential factor 
for the amplitude of tunneling wave. Accordingly, $e^{-S_b/\hbar}$ is 
the exponential factor for the current density of the tunneling wave,
which is directly related to the rate of decay. This is reflected
in Eq. (\ref{quantum-rate}). We want to remark that for one-dimensional
quantum mechanics, the tunneling rate (\ref{quantum-rate}) derived from
functional integral agrees with that obtained by standard WKB method of 
wave mechanics \cite{coleman}.


\begin{thebibliography}{10}
\bibitem{charging-energy}
The scale of the imaginary-time action, $\sqrt{E_J/E_C}$, can be
written as $E_J/\hbar\omega_p$. For small bias current $I\ll I_c$
($x\ll 1$), the ground-state energy ($\approx \hbar\omega_p/2$) must
be small compared to the barrier height $\Delta U$ ($\approx 2E_J$).
Otherwise, the system loses metastability. For $\sqrt{E_J/E_C}=2.18$
in our calculations, the barrier height is about $9$ times 
the ground-state energy and metastability can be safely assumed.
Further reduction of the junction capacitance would eventually
destroy this metastability.
\end{thebibliography}
\end{document}